# Long aging time thermal degradation of the ac conductivity and complex permittivity of conducting polypyrrole

I. Sakellis,  A.N. Papathanassiou and J. Grammatikakis

*University of Athens, Physics Department, Section of Solid State Physics, Panepistimiopolis, 15784 Zografos, Athens, Greece*

## Abstract

The modification of the ac conductivity and the complex permittivity of conducting polypyrrole was monitored throughout a two years thermal aging at 343K. Reduction of the cross-over frequency is correlated with the degradation of dc-conductivity, while the ac conductivity region corresponding to the so-called 'universal' dielectric relaxation remains practically invariant during the first year of aeging , which implies a collective co-operativity among multiple degradation processes that yield a practically time-independent effective disordered environment. A broad dielectric loss peak recorded in fresh specimens splits into two distinct relaxations for intermediate stages of the annealing process. The aging-time evolution of the dc component and the relaxations are qualitatively analysed and time constants are determined.

# 1. Introduction

Electric conductivity phenomena in conducting polypyrrole are the subject of numerous theoretical and experimental works, so as to have a unified picture of charge transfer in these materials [1, 2]. Conducting polymers are widely used as novel material used in technological applications such as solar cells, pH electrodes, media for hydrogen storage, electronic devices etc [3, 4, 5]. Electric charge transport in conducting polymers proceeds mainly by intra-chain and inter-chain hopping. This picture can be generalized, by considering highly conducting clusters of chains, forming conducting grains embedded in an insulating environment [6]. Conductive grains consist of polymer chains rich in polarons [7]. To a first approximation, metallic conduction occurs within the grain and hopping from one grain to another proceeds by phonon assisted tunneling through the insulating separation. The study of the thermal degradation of the electrical conductivity has been used to determine the stability of conducting polymers and blends, a crucial factor for their use in practical applications. In conductive polypyrrole, which has the granular metal structure, aging is attributed to the decreasing of the conducting grain size with a kinetic resembling a corrosion mechanism [8, 9, 3]. Moreover, the thermal degradation law serves as a tool to check if the structure of the conducting polymer is homogeneous or heterogeneous, of the granular metal type [9].

Conducting polypyrrole was aged at 343 K (isothermal aging or anneal) at ambient atmosphere for time duration of 2 years. The thermal degradation of the dc-conductivity $\sigma_0$ for short annealing duration ($t_a$) can be found in the literature (for example, see Ref. [8]). However, ac conductivity studies are more informative, since they provide information for the dynamics of electric charge carriers at different time scales (inversely proportional to the working frequency) and subsequently different length scales: Dc conductivity measurements can merely trace transport along the volume of the specimen through the percolation network, tracing the higher effective potential barriers of a disordered environment; ac measurements, apart from tracing the dc-conductivity constituents at low frequencies, are capable of detecting spatially localized charge transport and profile the entire potential energy landscape. Thermal degradation of the ac response of conducting polypyrrole was reported earlier by our group [11], but was restricted to short annealing time duration (up to 7 days) at aging temperature 343 K. Furthermore, the effect of pressure on controlled parallel sample physically aged (at room temperature) was studied [10] and the results were interpreted through the so-called cBΩ thermodynamic elastic model [12, 13, 14]. In principle, the time evolution of the conductivity induced by thermal annealing consists probably of different sequent stages. This is the reason of long-term thermal anneal studies.

## 2. Experimental Details

The sample preparation, which was reported earlier [9], consists of polymerizing freshly distilled pyrrole (Merc AR) in the presence of $FeCl_3$ as oxidant in hydrochloric acid–water solutions at pH 2.00 in an ice bath. The molar ratio of oxidant to monomer was 1:1 and the solvent used was triply distilled water. Polypyrrole was obtained as black powder and was purified by Soxhlet extraction for 36 h. Polypyrrole disc shaped specimens 13 mm in diameter and about 1.5 mm thick were made in a IR press.

The dielectric measurements were performed by placing the specimens in a sample holder of a vacuum cryostat operating at 1 Pa. The measurements were performed in the frequency range $10^{-2}$ to $2 \times 10^6$ Hz by a Solartron SI 1260 impedance analyzer. Two years keeping the sample inside an oven thermostated at 343 K in atmospheric conditions performed thermal aging. At certain times during the aforementioned 2 years anneal, the specimen was removed from the furnace and placed in the sample holder of the complex impedance analyzer system so as to record the complex conductivity and complex permittivity vs frequency at room temperature. After each dielectric measurement, which lasted for about 1 hour, the specimen returned inside the oven (at 343 K). In this way, the time evolution of the dielectric properties of conducting polypyrrole was sampled at certain intervals within the total two-years aging process.

## 3. Results and discussion

### *3.1 Effect of aging on the real part of the ac conductivity*

The complex conductivity spectrum as a function of frequency was measured at room temperature at certain time after the initiation of the thermal anneal, in order to monitor the effect of aging on the ac conductivity response of the material. Real part of the conductivity σ´ vs frequency f plots recorded for various aging intervals are shown in Fig. 1. These plots consist of a low-frequency plateau (dc-regime) and, above a critical crossover frequency $f_0$, a dispersive high-frequency region. When the frequency of the external a.c. field becomes larger than a percolation value, better use of the sites separated by lower potential barriers is made, the conductivity becomes dispersive on frequency and a dielectric loss mechanism takes place [15]. The effect of aging on ac conductivity is summarized as follows:

(i) The dc conductivity $σ_0$ gets reduced

(ii) The crossover frequency $f_0$ (highlighting the onset of the dispersive ac conductivity region) decreases

(iii) The measured (real part) of the ac conductivity σ´(f) is practically insensitive over a broad range of frequencies and ageing times of the dispersive region (i.e., conductivity plots obtained after different aging durations, almost collapse on a master curve, which can be described by a straight line of slope ~1 (in a log-log representation (Fig2)).

(iv) A high-frequency plateau shifts gradually to lower frequencies and the corresponding ac conductivity values get decreased.

The thermal degradation of the dc conductivity $σ_0$ of conductive polymers on aging duration is qualitative by the following law [16, 9]:

$$\sigma_0(t_a) = \sigma_0(t_a = 0) \cdot \exp\left(-\sqrt{\frac{t_a}{\tau^{aging}}}\right) \quad (1)$$

where $t_a$ denotes the aging time, $σ_0(t=0)$ is the dc conductivity of the fresh specimen and $\tau^{aging}$ is a time constant. The $\ln \sigma_0 \propto \sqrt{t}$ dependence is consistent with the granular metal structure, with conducting grains embedded into an insulating matrix, contracting with aging in a corrosion–like manner. Eq. (1) is well fitted to the $\sigma_0(t_a)$ data for the time duration $t_a$ from 100 to 300 days, which corresponds to an intermediate stage compared to the total duration (2 years). For longer aging times, $\sigma_0$ converges to a saturation value. In the inter-mediate aging region, the crossover frequency $\log f_0(t_a)$ is proportional to $\log \sigma_0(t_a)$ (Fig. 3). By fitting a straight line to the inter-mediate region data points we get $\sigma_0(t_a) \propto f_0(t_a)^{0.86}$. If a power law is assumed to match roughly the ac conductivity values *of the dispersive region* vs f, i.e., $\sigma(f) = Af^n$, where A and n are fitting parameters, we proved in a recent publication that in conducting polypyrrole the following equation holds [17, 18]:

$$\frac{\log A}{n} = -\log f_0 + \frac{\log \sigma_0}{n} \qquad (2)$$

The latter is valid for each one of the aforementioned conductivity curves, as can be seen in Fig.2: the dispersive region for inter-mediate aging time is almost insensitive to aging, thus, $\log A(t_a)/n(t_a)$ is constant. Subsequently, differentiating Eq. (2) with respect to the aging time $t_a$, we get $\sigma_0(t_a) \propto f_0(t_a)^n$. This relation correlates the exponent 0.86 derived from the $\log f_0$ vs $\log \sigma_0$ plot (Fig.3) with the power exponent n. The slope n of the log $\sigma(f)$ plots - estimated roughly 1 (Fig.2) - deviates about 15% the exponent value 0.86 derived from the $\log \sigma_0(\log f_0)$ plot (Fig.3), a fact that seems reasonable, since the power law fitting of ac conductivity data is a simple approximate manner to handle the ac response.

The fact that the dispersive region of the $\sigma'(f)$ plot for $t \leq 300$ days– which is a characteristic of disordered systems and is governed by the type of disorder (e. g., from the shape and width of the distribution of the length of conduction pathways accessible to the charge carriers [19]) - is not practically modified over such a long thermal aging (Fig.2) is quite remarkable. It seems that many different processes occur during the annealing process, such as the reduction of the size of conductive grains, polymer-chain complexes divide into smaller pieces etc, exhibit a collective co-operative behavior that yield an *effective* disordered environment which is insensitive to thermal aging.

*3.2 Effect of thermal aging on the dielectric loss spectrum*

In the formalism of the complex permittivity, the transition of the dc plateau to dispersive $\sigma$ as frequency increases results in the appearance of a relaxation peak. The imaginary part of the permittivity is $\varepsilon''(\omega) = \frac{\sigma'(\omega)}{\varepsilon_0 \omega} = \frac{\sigma_0}{\varepsilon_0 \omega} + \varepsilon_d''$, where $\varepsilon_0$ is the permittivity of free space and $\varepsilon_d''$ is the imaginary part of the permittivity after subtracting the dc component; i.e., $\varepsilon_d''(f) = [\sigma'(f) - \sigma_0]/2\pi f$). In the complex impedance representation, fresh polypyrrole and aged one for short time periods (up to 10 days) exhibit a unique broad dielectric loss mechanism [10]. In the time interval $t_a$ from 100 days to 300 days the above mentioned mechanism splits in a couple of relaxations, each one maximizing at a frequency $f_{max}$. Typical spectra of the imaginary part of the (relative) complex permittivity, after subtraction of the dc-component,(i.e., $\varepsilon_d''(f) = [\sigma'(f) - \sigma_0]/2\pi f$) are depicted in Fig. 4. The splitting of the broad dielectric loss mechanism recorded in fresh semi-conducting polypyrrole revealed two district peaks by applying pressure [20, 21]: one of them was insensitive to pressure, in contrast to the other one and were attributed to intra-chain (or intra-grain) and inter-chain (or inter-grain) electric charge flow, respectively. It seems that inter-mediate thermal anneal (for aging time from 100 to 300 days) is efficient – complementary to earlier dielectric loss under pressure experiments –to separate the aforementioned couple of mechanisms. The *effective* conductivity $\sigma^*$ calculated from the dielectric data is defined as: $\sigma_0^* = 2\pi f_{max} \varepsilon_0 \Delta\varepsilon$, where $\varepsilon_0$ is the permittivity of free space and $\Delta\varepsilon$ is the strength of the relaxation mechanism. $\ln\sigma_0^*(t_a)$ seems to be proportional to $\sqrt{t_a}$ so, Eq. (1) is also used to fit the $\sigma_0^*(t_a)$ data in order to estimate the aging time constants for each dielectric loss peaks (Fig.5). The aging time constants for the dc conductivity is $\tau^{aging} = (1.9 \pm 0.2)$ days, while, for the effective conductivity $\sigma^*$ (calculated from $f_{max}$ and $\Delta\varepsilon$ for each loss peak) is $\tau^{aging} = (1.7 \pm 0.2)$ days for each of the relaxation mechanisms, respectively. For time duration larger than 300 days both the dc-conductivity and the effective conductivity $\sigma^*$ practically saturates. The square root of time dependence probably indicates a diffusion like mechanism related to the reduction of conducting regions [27].

## 3.3. Effect of thermal aging on the applicability of the Barton–Nakajima–Namikawa (BNN) model

dc and ac conductivity of many disordered materials are interconnected through the Barton–Nakajima–Namikawa (BNN) model [22, 23], which implies that the dc conductivity $\sigma_0$ is correlated with the frequency $f_{max}$ where dielectric loss has its maximum and its dielectric strength $\Delta\varepsilon$, through:

$$\sigma_0 = p 2\pi f_{max} \varepsilon_0 \Delta\varepsilon \qquad (3)$$

where p is an empirical constant, close to 1. Such correlation in conductive polymers [24, 25, 26] is justified in some cases but fails in others. Divergences from the BNN model, that was observed in some amorphous materials, was attributed to over-simplifications of the theoretical model; the material characteristics and the nature of charge carriers are ignored, a fixed hopping mechanism is assumed, dipole relaxation and many body long range interaction are ignored etc. [24]. Keeping the notation $\sigma_0^* = 2\pi f_{max} \varepsilon_0 \Delta\varepsilon$ of section 3.2, Eq. 30 can alternatively appear as:

$$\sigma_0 = p\sigma_0^* \qquad (4)$$

For the dielectric relaxation peaks detected in the aging experiment, $\sigma_0^*$ is plotted against the measured dc conductivity $\sigma_0$ in Fig. 6. We observe that for intermediate aging times (from 100 to 300 days) the data points scatter around the prediction of the BNN model, for p=1. This indicates that the parameter p is not an aging invariant parameter, but changes during the aging process. This observation is in accordance with an earlier publication of ours [10], where $\sigma_0$ and $\sigma_0^*$ exhibited different aging time constants (for short aging duration up to 7 days), which indicated a variation of the parameter p with aging.

## 5. Conclusion

As a conclusion, the investigation of the thermal degradation – over 2 years total duration – of the ac conductivity and complex permittivity of conducting polypyrrole reveals interesting features: The dc conductivity correlates with the crossover critical frequency. A remarkable aging-insensitive dispersive region of the ac conductivity is observed; its power-law dependence upon frequency is compatible with predictions with a recent phenomenological model of ours [17, 18]. The aging-insensitive dispersive region indicates probably that many different processes occur during the annealing process (such as the reduction of the size of conductive grains, polymer-chain complexes divide into smaller pieces etc), exhibit a collective co-operative behavior that yields an *effective* disordered environment which is insensitive to thermal aging. A broad dielectric loss peak observed in fresh specimen, splits in a couple of distinct relaxation mechanisms for intermediate aging time (from 100 to 300 days). The effective conductivity estimated from the characteristics of these dielectric responses, as well as the dc conductivity, are square root functions of the aging time with comparable aging time constants. The empirical proportionality parameter appearing in the BNN model proved to be aging-dependent, in agreement with previous results obtained frm short-time aging.

**Figure Captions**

Figure 1: The (real part) of the ac conductivity vs frequency measured at room temperature after different duration aging times at 343K. From top to bottom: 0, 34, 138, 191, 260, 329, 487, 515, 689 and 773 aging time in days, respectively. The thick straight line has slope 1 and was drawn to help the reader.

Figure 2: The (real part) of the ac conductivity vs frequency measured at room temperature after different duration aging times at 343K. From top to bottom: 0, 34, 138, 191, 260, 329, aging time in days, respectively.

Figure 3: The crossover frequency $f_0$ vs dc conductivity $\sigma_0$ for various aging times. The straight line is the best fit to the data points.

Figure 4: Typical dielectric loss spectra recorded) in fresh, and aged polypyrrole for 191, 260, 329 and 487 days.

Figure 5: The dc conductivity $\sigma_0$ (stars) and the effective conductivity $\sigma^*$ of the low (squares) and high (triangles) frequency dielectric loss peak, respectively, against the square root of the aging time. Straight lines are best fits to the data points for aging time from 100 to 300 days.

Figure 6: The effective dc conductivity $\sigma^*_0$ calculated from the complex permittivity data ($f_{max}$ and $\Delta\varepsilon$) for the low frequency peak (squares) and the high frequency one (triangles) as a function of the measured dc conductivity $\sigma_0$, determined from the low frequency limit of $\sigma'(f)$. The solid line is the prediction of the BNN model (Eq. (4)) for its (empirical) parameter p=1.

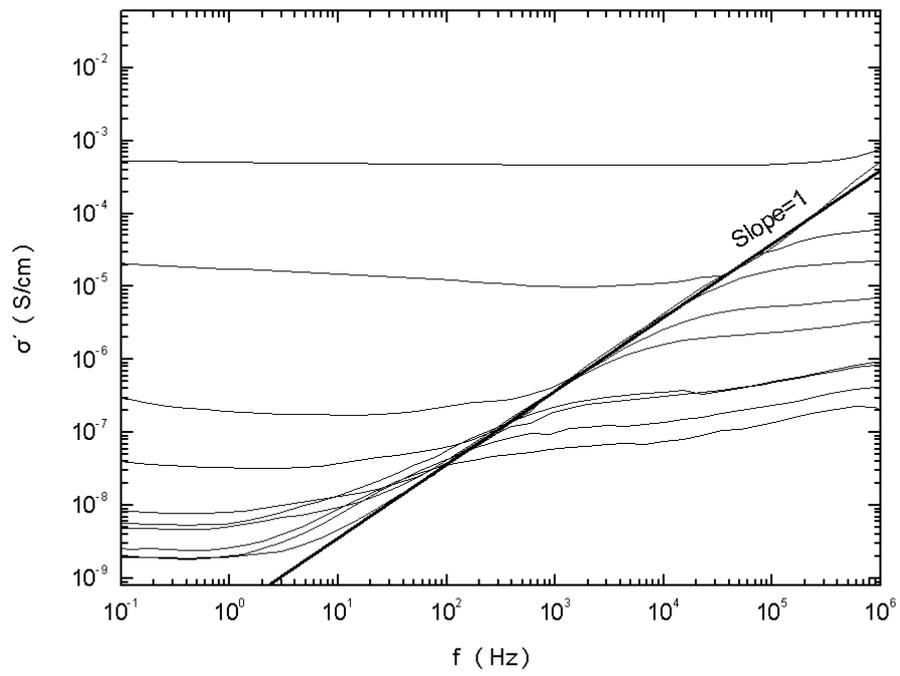

FIGURE 1

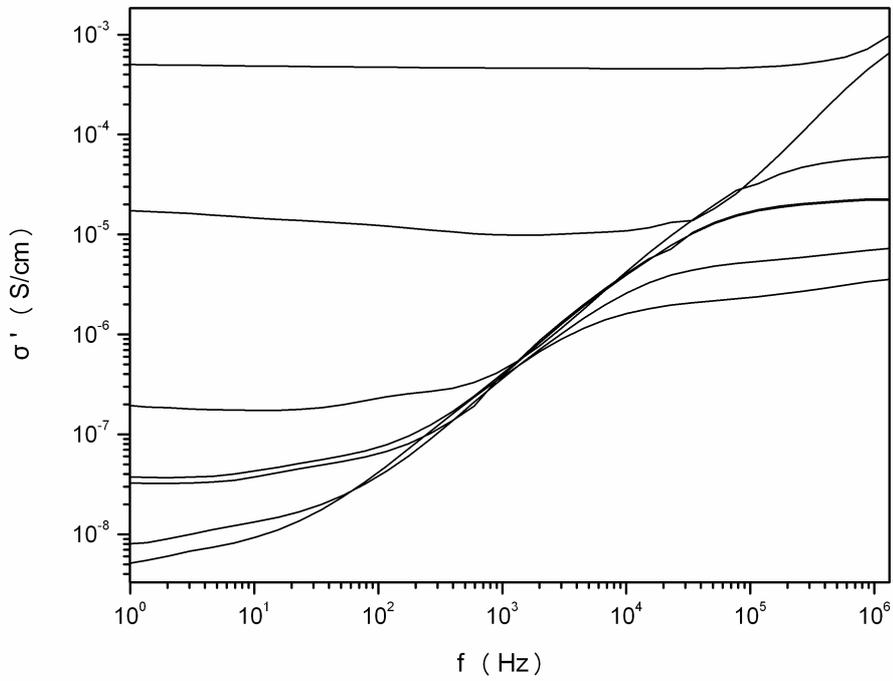

FIGURE 2

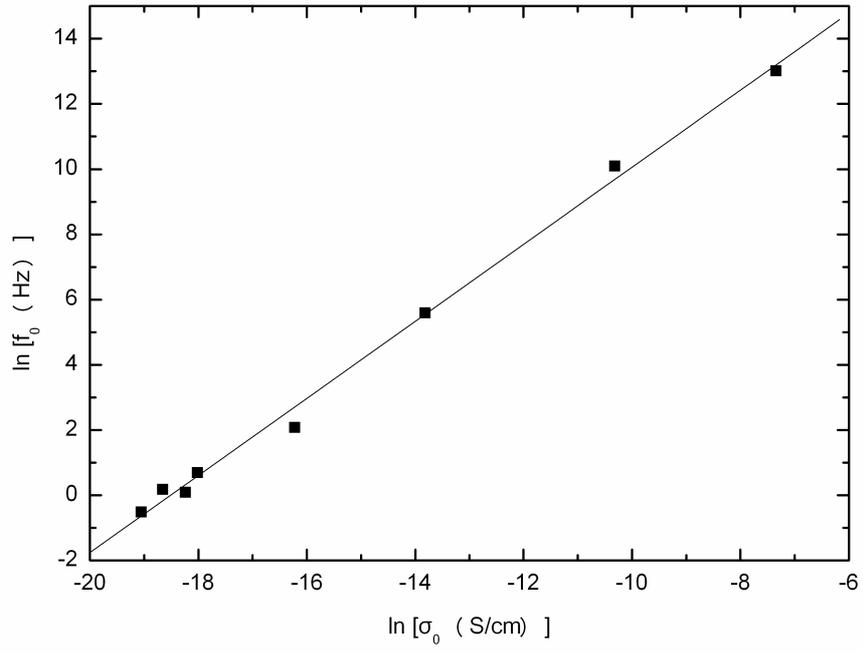

FIGURE 3

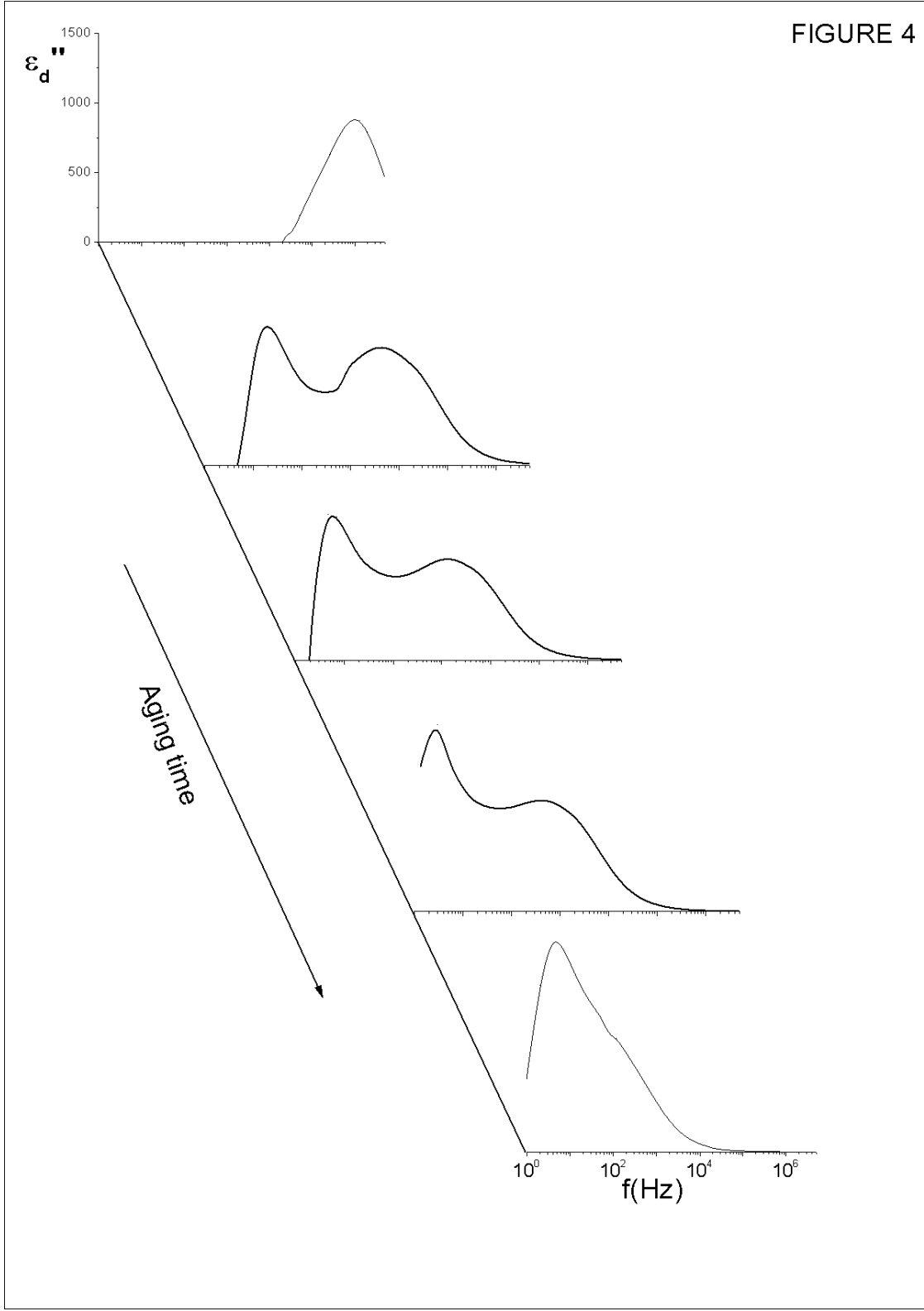

FIGURE 4

FIGURE 5

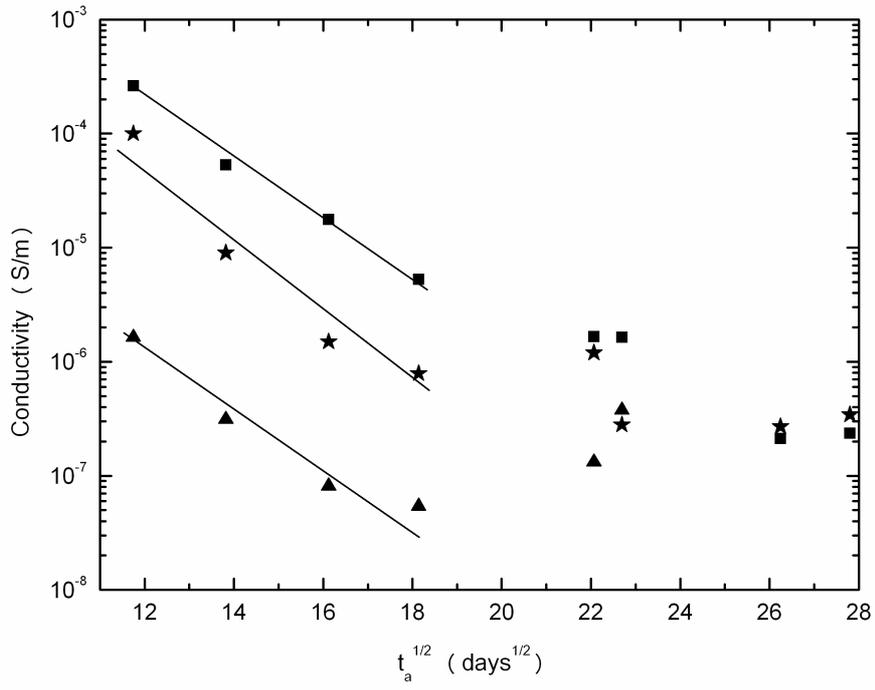

FIGURE 6

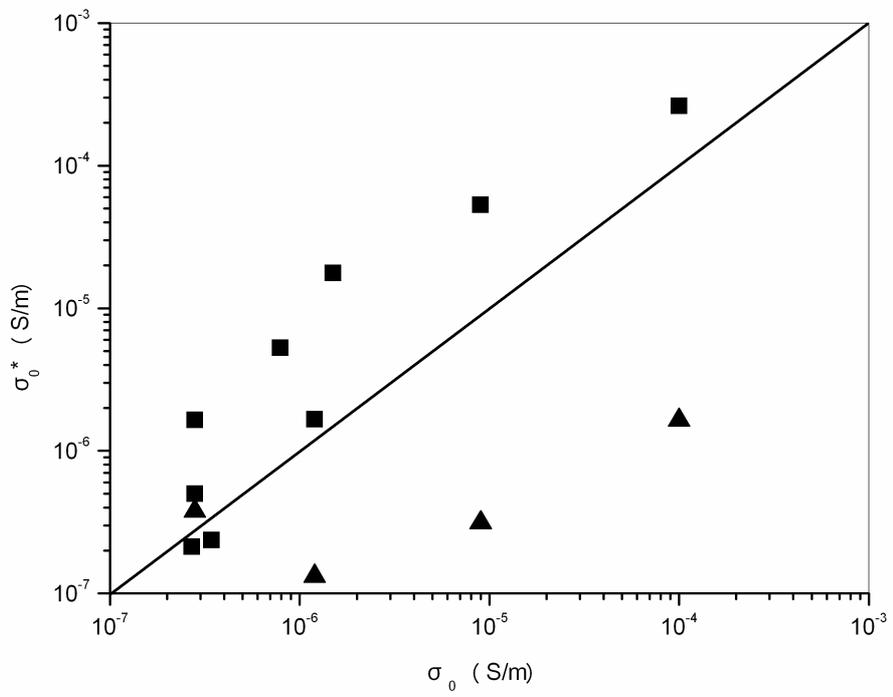